\documentclass[twocolumn]{aastex63}
\usepackage{graphics,epsf}
\usepackage{amsmath}                
\usepackage{amsfonts}               
\usepackage{amssymb}                
\usepackage{epsfig}                 
\usepackage{appendix}
\usepackage{graphicx}
\usepackage{float}
\usepackage{color}
\usepackage{multirow}
\usepackage{colortbl}
\usepackage[para,online,flushleft]{threeparttable}

\hypersetup{citecolor=blue, 
            linkcolor=red, 
            menucolor=blue, 
            urlcolor=blue}  

\newcommand{\cm}{{~\rm cm}}
\newcommand{\m}{{~\rm m}}
\newcommand{\km}{{~\rm km}}
\newcommand{\s}{{~\rm s}}
\newcommand{\h}{{~\rm h}}

\newcommand{\g}{{~\rm g}}
\newcommand{\G}{{~\rm G}}
\newcommand{\K}{{~\rm K}}
\newcommand{\erg}{{~\rm erg}}

\newcommand{\days}{{~\rm days}}

\begin{document}

\title{Pre-explosion, explosion, and post-explosion jets in supernova SN 2019zrk}


\author[0000-0003-0375-8987]{Noam Soker}
\affiliation{Department of Physics, Technion, Haifa, 3200003, Israel;  soker@physics.technion.ac.il}

\begin{abstract}
I analyse some properties of the luminous transient event SN~2019zrk and conclude that jets were the main powering sources of the pre-explosion outburst (pre-cursor) and ejection of a massive circumstellar matter (CSM), of the very energetic explosion itself, and of the post-explosion bump in the light curve.  The pre-explosion energy source is mainly a companion (main sequence, Wolf-Rayet, neutron star or black hole) star that accreted mass and launched jets. I find that the fast expansion of the CSM after acceleration by the explosion ejecta requires the explosion energy to be $\ga 10^{52} \erg$. Only jet-driven explosions can supply this energy in such SN~2009ip-like transients. I conclude that ejecta-CSM interaction is extremely unlikely to power the bright bump at about 110 days after explosion. Instead, I show by applying a jet-driven bump toy-model that post-explosion jets are the most likely explanation for the bump. I leave open the question of whether the explosion itself (main outburst) was a core collapse supernova (CCSN) or a common envelope jets supernova (CEJSN). In this study I further connect peculiar transient events, here 2009ip-like transient events, to CCSNe by arguing that jets drive all events, from regular CCSNe through superluminous CCSNe and to many other peculiar and super-energetic transient events, including CEJSNe. Jet-powering cannot be ignored when analyzing all these types of transients.  
 \end{abstract}


\section{Introduction} 
\label{sec:intro}

The SN~2009ip-like (2009ip-like) class includes several transients that show prominent pre-cursor with a light curve that has relatively fast rise and decline to peak luminosity, bumps during the decline phase, and cannot be explained by regular core collapse supernovae (CCSNe). Either there are extra ingredients in a CCSN scenario, or they are descendants of different evolutionary routes. In addition to SN~2009ip (e.g., \citealt{Smithetal2010, Fraseretal2013, Mauerhanetal2013, Pastorelloetal2013, Grahametal2014, Smithetal2022}), some other class members that show (most of) these properties are SN~2010mc (e.g., \citealt{Ofeketal2013, Smithetal2014}), SN~2013gc \citep{Reguittietal2019}, LSQ13zm \citep{Tartagliaetal2016}, SN~2015bh (e.g., \citealt{EliasRosaetal2016, Thoneetal2017}),  SN~2016bdu (e.g., \citealt{Pastorelloetal2018}), and AT~2016jbu (e.g., \citealt{Boseetal2017ATel, Kilpatricketal2018, Brennanetal2022a}). 

In this study I will refer to the main luminosity peak as `explosion', to the pre-cursor as the pre-explosion phase, and to the late decline phase of the light curve as the post-explosion phase. I note though that the `explosion' is not necessarily a terminal explosion that ends the evolution, e.g., of a binary system, neither it implies the collapse of the core, although this might be the case.  

Most researchers agree that these events involve a massive circumstellar matter (CSM). However, there is no consensus on some basic ingredients of 2009ip-like events. One issue in disagreement is the powering of the main events. One view (e.g., \citealt{Smithetal2014} for SN~2009ip) is that, at least in come cases, the precursor (pre-explosion outburst) is a CCSN event and that the main peak of the light curve (the `explosion') results from ejecta-CSM interaction.   
Another class of possibilities is that the explosion is a CCSN while the pre-explosion activity results from either an instability of the CCSN progenitor or from a binary interaction (section \ref{sec:PreExplosion}). Another model for SN~2009ip was the merger of a massive star with a luminous blue variable (LBV) star \citep{SokerKashi2013}, and much of the energy results from accretion onto the main sequence star. Then there is the possibility that the explosion is driven by the accretion power onto a neutron star (NS) or onto a black hole (BH) companions that spiral-in in a common envelope evolution (CEE) inside the RSG envelope (e.g., \citealt{Gilkisetal2019}), or deep to the RSG core (e.g., \citealt{Chevalier2012, Schroderetal2020})

Another issue is the role that jets play in 2009ip-like events. Jets can be launched by the newly born NS or BH at the center of the collapsing core, or by a NS/BH that experience a CEE with a RSG and/or its core. Transient events that are powered mainly by jets that a NS/BH launches inside the envelope or the core of a RSG are termed common envelope jets supernovae (CEJSNe; for some recent studies that emphasise the role of jets see, e.g., \citealt{Gilkisetal2019, SokeretalGG2019, GrichenerSoker2019a, LopezCamaraetal2019, LopezCamaraetal2020MN, AkashiSoker2021, GrichenerSoker2021, Gricheneretal2021, Schreieretal2021, Soker2021, Hilleletal2022}, and for studies of NS/BH merger with the RSG core that do not emphasize jets see, e.g., \citealt{FryerWoosley1998}; \citealt{Chevalier2012}; \citealt{Schroderetal2020}). Strictly speaking, if the NS/BH does not enter the core the event is a CEJSN-impostor, but in this study I will use CEJSN to refer to impostors as well. 

The diversity of outflow properties and light-curves that CEJSNe allow (see above papers) have made them attractive scenarios for some rare peculiar explosions. These include among others the peculiar gamma ray burst (GRB) 101225A \citep{Thoneetal2011}, iPTF14hls \citep{SokerGilkis2018}, AT2018cow and other fast blue optical transients \citep{SokeretalGG2019, Metzger2022FBOT, Soker2022FBOT}, and the luminous radio transient event VT~J121001+495647 \citep{Dongetal2021}. 
It is possible that (some) 2009ip-like events are also CEJSNe (e.g., \citealt{Gilkisetal2019, Schroderetal2020}). In any case, I argue that jets seem to play a crucial role in some or all phases of 2009ip-like events.  
 
In this study I argue that jets must play the major roles in the pre-explosion (section \ref{sec:PreExplosion}), in the explosion (section \ref{sec:Explosion}), and in the post-explosion (section \ref{sec:PostExplosion}) powering of the 2009ip-like event SN~2019zrk. 

\cite{Franssonetal2022} present a thorough observational study of SN~2019zrk (for an early study see \citealt{Strotjohanntal2021}). It reached a maximum magnitude of $M_{\rm r}=-19.2$ and showed expansion velocities of up to $v_{\rm ex} \simeq 16,000 \km \s^{-1}$. Its light curve is complex, with precursor (pre-explosion emission) and bumps in its light curve. I will analyse and model the sharp and relatively bright bump centred at $t=110~{\rm d}$ in section \ref{sec:PostExplosion}. The total radiate energy during the main outburst (which I also refer to as explosion) is $E_{\rm rad,ex} \simeq 5 \times 10^{49} \erg$. \cite{Franssonetal2022} point out that like the case with SN~2009ip \citep{Marguttietal2014} the total radiated energy can be much larger due to substantial contribution from UV and X-rays.
   
By its peak magnitude of  $M_{\rm r} =-19.2$ SN~2019zrk is a luminous supernovae (see definition by \citealt{Gomezetal2022}). In earlier papers I argued that luminous supernovae \citep{Soker2022LSNe}, and more so super superluminous supernovae which have $M_{\rm r} < -20$ \citep{SokerGilkis2017}, are most likely powered by jets, even if an energetic magnetar supply energy as well (e.g., \citealt{Soker2016Magnetar, Soker2017Magnetar}).  \cite{Reichertetal2022} conduct magnetohydrodynamic simulations of jets in CCSNe and further support jet-powering of superluminous CCSNe. 
This discussion further motivates me to consider jet-powering of SN~2019zrk. In my summary (section \ref{sec:Summary}) I will connect 2009ip-like events to other luminous and superluminous CCSNe.  

\section{Pre-explosion jets and merger-driven mass loss} 
\label{sec:PreExplosion}

Like other systems in the 2009ip-like class, SN~2019zrk also displays an energetic pre-cursor, i.e., pre-explosion outburst(s). Such outbursts have some similarities with other types of outbursts of the heterogeneous class of intermediate luminosity optical transients (ILOTs; other names include gap objects; luminous red novae; intermediate luminosity red transients), e.g., as \cite{Smithetal2010} and \cite{SokerKashi2013} discussed for SN~2009ip and \cite{Brennanetal2022b} discussed for AT~2016jbu.  
  
\cite{Franssonetal2022} find the merger scenario to  be the most promising to explain the properties of SN~2019zrk. It is not clear whether they refer to a main sequence companion that enters the envelope of a RSG as \cite{SokerKashi2013} and \cite{Kashietal2013} discussed for SN~2009ip, or whether they refer to a NS (or even a BH) companion that enters the RSG envelop as \cite{Gilkisetal2019} suggested already for SN2009ip. 
In both cases \textit{jets play critical roles} as these papers pointed out. 

Let me elaborate on this by referring to two problems that \cite{Franssonetal2022} mention in their discussion of SN~2019zrk. Both of these have already been solved with jets. 

\cite{Franssonetal2022} claim that a possible problem for the merger model might be the large velocity observed  in the precursor of SN 2009ip, $\approx 12,000 \km \s^{-1}$ (\citealt{Foleyetal2011, Pastorelloetal2013}). I disagree, because even for a main sequence companion that launches jets at $\approx 2000-3000 \km \s^{-1}$ the interaction of the jets with the CSM can accelerate gas to velocities of $\ga 10,000 \km \s^{-1}$, as \cite{TsebrenkoSoker2013} showed for parameters that fit SN~2009ip and \cite{AkashiKashi2020} showed for the parameters that fit the Great Eruption of Eta Carinae (an energetic ILOT event). 
For a NS/BH companion that might lunch very fast jets, as in the CEJSN scenario, this is not a problem at all, as \cite{Gilkisetal2019} argued for SN~2009ip. 

The second problem that \cite{Franssonetal2022} mention relates to the formation of a massive CSM just before the explosion. \cite{Franssonetal2022} write that the energy that core-convection-driven waves (e.g., \citealt{QuataertShiode2012, WuFuller2021}) carry to the envelope of massive RSG stars is too low to explain the ejection of a pre-explosion massive CSM. \cite{McleySoker2014} already solved this problem. \cite{McleySoker2014} found indeed that the main outcome of energy that the waves deposit to the RSG envelope is envelope expansion rather than mass ejection. They concluded that pre-explosion outbursts result from the accretion energy that a binary companion releases as it accretes mass from the extended envelope. \cite{DanieliSoker2019} studied this process for a NS companion that accretes mass from the inflated envelope and launches jets.  

\cite{Franssonetal2022} conclude that ". . . we believe that the low mass alternative, involving a merger, is the one with the least problems." I accept this, and add that jets constitute the main powering of the outflow. 
Of course, the energy source is the accretion onto the NS companion, as with all CEJSNe.
This basic idea goes back to \cite{SokerKashi2013} and \cite{Kashietal2013} who discussed merger with a main sequence companion (also \citealt{SokerKashi2016} for accretion and jets in SN~2015bh). However, for the more luminous SN~2019zrk I prefer the CEJSN scenario, where the companion that enters a CEE with the RSG is a NS as \cite{Gilkisetal2019} suggested already for SN2009ip. \cite{Schroderetal2020} consider the CEJSN scenario for SN~1998S, which is a 09-like SN. 

\begin{figure*}
 \centering
\includegraphics[trim= 1.7cm 0.1cm 0.0cm 3.0cm,scale=0.80]{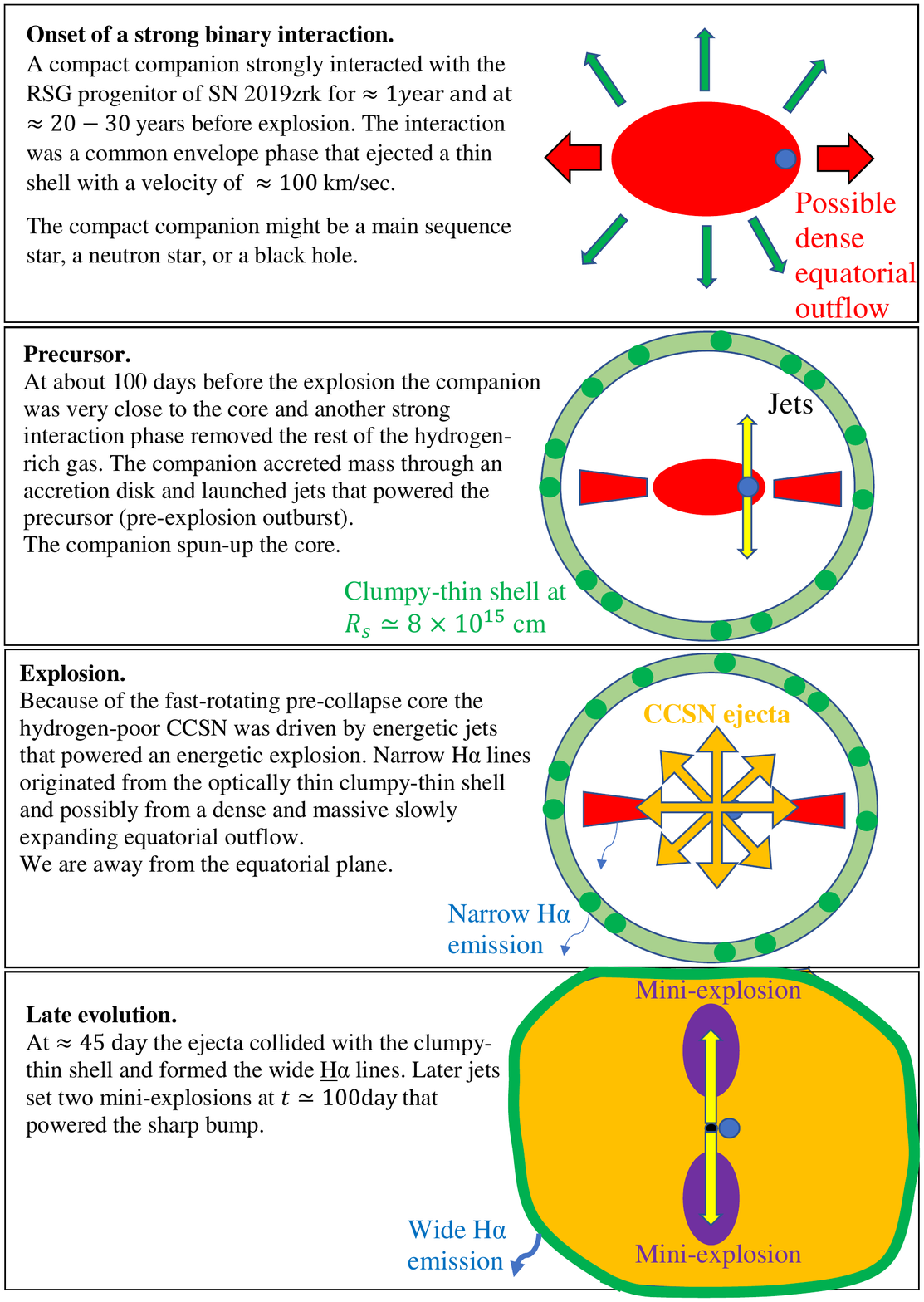}
\caption{A schematic description (not to scale) of the scenario that I propose in this study where the final explosion is a CCSN. Similar energetic transient events might be powered by CEJSNe, namely, where a NS/BH companion enters the core, destroys the core and launches energetic jets.    
}
 \label{Fig:schematic}
 \end{figure*}

My main point in this section is that jets powered the pre-explosion outburst (pre-cursor) of SN~2019zrk, as I schematically show in the second panel of Fig. \ref{Fig:schematic}. I suggest that this outburst was preceded by a longer CEE phase (first panel of Fig. \ref{Fig:schematic}).
   
\section{Explosion jets} 
\label{sec:Explosion}

In this section I show that both the CSM model that \cite{Franssonetal2022}) consider and an alternative CSM model that I discuss here require that jets power the explosion (main outburst) of SN~2019zrk.

The spectra of SN~2019zrk (figure 6 of \citealt{Franssonetal2022}) at early times $t \la 45~{\rm day}$ are composed from a blue continuum with superimposed narrow emission lines with faint broad wings that \cite{Franssonetal2022} attribute to electron scattering. At $t \simeq 45~{\rm day}$ and later the spectra show a broad and asymmetric line profile of H$\alpha$. 
\cite{Franssonetal2022} conclude that the H$\alpha$ emission is decoupled from the thermalization photosphere. 

To explain the transition from narrow to broad H$\alpha$ emission lines at $t\simeq 45~$days \cite{Franssonetal2022} build a flow model where the CSM is optically thick during the early time period. They then estimate the CSM mass in front of the shock at 50 days to be $M_{\rm CSM,F22}=0.96 M_\odot$, where `F22' marks quantities as \cite{Franssonetal2022} deduce. Such a CSM mass has a severe demand on the explosion energy. Because the line widths even at $t=110$ days show a velocity of $v_{\rm ex}\simeq 16,000 \km \s^{-1}$, the explosion ejecta must accelerate the hydrogen-rich shell to these velocities. Since very little energy relative to the explosion energy is radiated away, I consider energy conservation. Consider an ejecta shell of mass $M_{\rm f}$ at the front of the ejecta moving with a velocity of $v_{\rm ej}$ and colliding with the shell of mass $M_{\rm CSM}$. The collision accelerates the CSM to velocity $v_{\rm ex}$. Energy conservation reads 
\begin{equation}
\frac{1}{2} M_{\rm f} v^2_{\rm ej}
\simeq\frac{1}{2} \left( M_{\rm f} + M_{\rm CSM} \right) v^2_{\rm ex}.
\label{eq:Econserv}
\end{equation}
The collision will take place at radius $v_{\rm ej}\times 50~{\rm days}$, the distance of the front of the ejecta, rather than at $v_{\rm ex}\times 50~{\rm days}$ as \cite{Franssonetal2022} take. The demand of optically thick CSM implies that the CSM mass now is $M_{\rm CSM}=M_{\rm CSM,F22}(v_{\rm ej}/v_{\rm ex})^2$.  Substituting this mass in equation (\ref{eq:Econserv}) and demanding minimum possible ejecta energy to obey energy conservation I find the ejecta velocity to be $v_{\rm ej} = 2^{1/2} v_{\rm ex}$, where here $v_{\rm ex}=16,000 \km s^{-1}$, and $M_{\rm f}= 2 M_{\rm CSM,F22}=1.92 M_\odot$. The constraint on the total energy of the ejecta under the condition of optically thick CSM is then 
\begin{equation}
\begin{aligned} 
E_{\rm ej,tot} > 2 \times 10^{52} 
\left( \frac{f_{\rm E,f}}{0.5} \right)^{-1}
\left( \frac{f_{\rm k}}{1} \right)^{-1} 
\left( \frac{M_{\rm CSM,F22}}{0.96 M_\odot} \right) 
\erg, 
\end{aligned}
\label{eq:Ef}
\end{equation}
where $f_{\rm E,f}$ is the fraction of the energy that the front of the ejecta that collides with the CSM has relative to the entire ejecta, and $f_{\rm k}$ is the fraction of energy that ends as kinetic energy rather than thermal energy during the collision.
This minimum energy requires the CSM mass to be twice as large as what \cite{Franssonetal2022} find. The energy is not likely to be at its minimum, and the total ejecta energy is likely to be much larger. 

The conclusion from this discussion is that the model that  \cite{Franssonetal2022} build for the CSM demands an extremely energetic explosion with an explosion energy of $E_{\rm ej,tot} \simeq {\rm several} \times 10^{52} \erg$. Only jets can supply this energy as the neutrino-driven mechanism is limited to explosions energies of $< 3 \times 10^{51} \erg$ (e.g. \citealt{Ertletal2016, Sukhboldetal2016}).  

Another problem is as follows. In many cases the collision of the ejecta with the CSM is applied to explain bright supernovae. But for the above parameters the collision channelled a negligible amount of energy to radiation. The total observed radiated energy at $t>50~{\rm days}$ is $<2\times 10^{49} \erg$.
The acceleration of the CSM requires a minimum energy of 
$0.5M_{\rm CSM,F22}v^2_{\rm ex}=2.4\times 10^{51} \erg$.
Namely, the very energetic collision channels $<1\%$ of the collision energy to radiation. This is a very small ratio that requires an explanation considering that the shell optical depth is $\approx 1$ at collision time in the model of \cite{Franssonetal2022} and that collision takes place at a very large radius (hence the gas suffers little adiabatic cooling in the following several weeks).   

As a plausible alternative, I propose a model with a much lower-mass CSM. I propose that the photosphere up to $t \simeq 45~{\rm days}$ was hydrogen-free. The narrow emission lines result from the CSM that has been ionised from the explosion time on, as in the model of \cite{Franssonetal2022}. But I do not require this CSM to be optically thick. Therefore, its mass might be much smaller, namely, I require that the mass that the ejecta collided with in the first $\simeq 110 \days$ is $M_{\rm CSM} \ll 1 M_\odot$. The CSM resides in a dense shell in the zone $r>v_{\rm ej} \times 50~{\rm days}$. The velocity of the ejecta can be larger than the maximum observed velocity, $v_{\rm ej} > v_{ex} = 16,000 \km \s^{-1}$, but after the ejecta-CSM collision the velocity is $v_{\rm ex} \simeq 16,000 \km \s^{-1}$. 
The hydrogen shell is sufficiently dense to have a strong H$\alpha$ emission. 

Let me give one example of a set of plausible physical values. 
Consider the front of the ejecta that collides with the CSM to expand with a velocity of $v_{\rm ej} \simeq 20,000 \km \s^{-1}$ and have a mass of $M_{\rm ej,front}$. 
Here I apply momentum conservation as I require that the CSM is accelerated in a very short time, before the shocked CSM and shocked ejecta have time to adiabatically cool and channel thermal energy to kinetic energy. Using energy conservation will require less ejecta energy even. 
To accelerate the CSM shell of mass $M_{\rm CSM}$ to $v_{\rm ex} = 16,000 \km \s^{-1}$ the requirement from momentum conservation is $M_{\rm ej,front}=4 M_{\rm CSM}$.
The energy that the front of the ejecta carry is 
$E_{\rm ej,front} \simeq 3.2 \times 10^{51} (M_{\rm CSM}/0.2M_\odot) \erg$. The energy that is dissipated to thermal energy during the collision is $E_{\rm dis}=0.2 E_{\rm ej,front}$.  

This very fast ejecta front that collides with the CSM might crudely carry fifth of the ejecta mass and a larger fraction of its energy. For example, for an ejecta density of $\rho_{\rm ej} \propto r^{-1}$ (e.g., \citealt{SuzukiMaeda2019}) the fraction of mass in the velocity range of $18,000 \km \s^{-1}$ to the maximum value of $v_{\rm ej}=20,000$ that I expect to collide with the CSM during the time period $t=40~{\rm days}$ (just before the time of the first spectrum with wide lines) to $t=80~{\rm days}$ (just before the time of the first spectrum with wide lines) carry a fraction of $0.19$ of the total ejecta mass. It carry a fraction of $\simeq 35\%$ of the ejecta energy. Overall, the constrain on the ejecta mass and energy under the assumption of $v_{\rm ej} \simeq 20,000 \km \s^{-1}$ and the observations of $v_{\rm ex} \simeq 16,000 \km s^{-1}$ are
\begin{equation}
\begin{aligned} 
M_{\rm ej} \approx  4  
\left(\frac{M_{\rm CSM}}{0.2 M_\odot} \right)
M_\odot,
\end{aligned}
\label{eq:Mejecta}
\end{equation}
and
\begin{equation}
\begin{aligned} 
E_{\rm ej} \approx  10^{52} 
\left(\frac{M_{\rm CSM}}{0.2 M_\odot} \right)
\erg,
\end{aligned}
\label{eq:Eejecta}
\end{equation}
respectively. 
These values are not unique. I only point out the possibility of an event of lower mass and energy than what the model of \cite{Franssonetal2022} requires. In section \ref{sec:PostExplosion} I will consider parameters that are between these two models. 

The maximum H$\alpha$ luminosity is $L_{\rm H \alpha}({\rm obs,m})\simeq 2 \times 10^{41} \erg \s^{-1}$ \citep{Franssonetal2022}. To account for this luminosity I need to constrain the properties of the hydrogen-rich shell that I place at a radius of 
$R_s \simeq 20,000 \km \s^{-1} \times 45 ~{\rm day}=7.8 \times 10^{15} \cm$, where the ejecta collides with it after $\simeq45~{\rm day}$. I scale the shell width with $\Delta r = 0.05R_s$, and the compression of the hydrogen-rich gas beyond the shock by $\beta = 7$. The expected maximum luminosity of this shell for Case B recombination at a temperature of $10^4 \K$ is 
\begin{eqnarray}
\begin{aligned} 
L_{\rm H\alpha} &  \simeq  3 \times  10^{41} 
\left(\frac{M_{\rm CSM}}{0.2 M_\odot} \right)^2
\\ & \times
\left(\frac{\Delta r}{0.05 R_s} \right)^{-1}
\left(\frac{\beta}{7} \right)
\erg \s^{-1}.
\end{aligned}
\label{eq:LHalpha}
\end{eqnarray}
Such a shell can account for the observed maximum H$\alpha$ luminosity. 
However, before the collision of the ejecta with the shell the shell is not compressed but the observed H$\alpha$ luminosity is already $L_{\rm H \alpha}({\rm obs})\simeq 1.5 \times 10^{41} \erg \s^{-1}$ \citep{Franssonetal2022}. Therefore, to account for the H$\alpha$ luminosity at early time as well the shell must by denser. Either it is more massive with $M_{\rm CSM} \simeq 0.4M_\odot$ and/or it is clumpy with dense clumps. Even for a mass of $M_{\rm CSM} \simeq 0.4M_\odot$ the optical depth of such a shell is $\tau \simeq 0.33$. Therefore, it is indeed optically thin. I prefer to consider a very clumpy shell with a mass of $M_{\rm CSM} \simeq 0.2- 0.3M_\odot$. Future hydrodynamical simulations that include radiative transfer should explore such a flow structure.

I also note that the H$\alpha$ to H$\beta$ luminosity ratio at early times is larger than a simple Case B recombination. This points to a more complicated flow structure. For example, there might be very low-mass ejecta at very high velocities, namely very large energy but very small momentum, that collides with the dense shell at very early times. This collision adds to the excitation of hydrogen in the shell but does not accelerate it much, in particular if the shell is composed of dense clumps. Another possibility is that there is a massive hydrogen-rich equatorial slow outflow, i.e., in the equatorial plane of the system (e.g., SN 1987A) that is more or less on the plane of the sky (see below) starting close to the star. Therefore the collision of the fast ejecta with this equatorial outflow takes place at early times. Part of the H$\alpha$ emission is due to the collision of the very fast (and optically thin) ejecta with this dense equatorial outflow.

I emphasize again that I differ from \cite{Franssonetal2022} mainly in that I do not demand the CSM to be optically thick in the first $\sim 50~{\rm days}$. This leads to a model with lower explosion energy and lower ejecta mass.  The question then is where is the heavy hydrogen-rich  envelope? In this scenario it is either concentrated in an equatorial dense outflow (as I mentioned above; see Fig. \ref{Fig:schematic}), or it was expelled at a much earlier time. Better, both effects. Namely,  some of the hydrogen-rich envelope was expelled as the NS interacted with the envelope before it entered a CEE. Only after thousands of years or more the NS entered a CEE with the RSG progenitor.    

The main conclusion of this section is that the explosion energy is $E_{\rm ej} \ga 10^{52}$, and that only jets can supply this explosion energy, whether in a CCSN event or in a CEJSN event. 

\section{Post-explosion jets} 
\label{sec:PostExplosion}
\subsection{The bump in the declining phase} 
\label{subsec:Bump}

In this section I consider the sharp bump that lasts from $\simeq t=95~{\rm days}$ and reach a peak luminosity at $t_{\rm bump} \simeq 110~{\rm days}$, and then declines on a similar time scale. I take the observed timescale of the bump to be $t_{\rm b,obs} = 15~{\rm  \days}$. \cite{Franssonetal2022} find the total extra radiated energy in this bump to be $E_{\rm rad,b}\simeq 1.8 \times 10^{48} \erg$. 
For a later application I take the `typical maximum luminosity' of the bump to be $L_{\rm b,obs} =  E_{\rm rad,b}/t_{\rm b,obs} \simeq 1.4 \times 10^{42} \erg \s^{-1}$. 

A key observation to this discussion is that the spectra at $t=83~{\rm days}$ (pre-bump) and at $t=109~{\rm days}$ (bump) are very similar.

\subsection{Limitations of CSM powering} 
\label{subsec:CSM}
\cite{Franssonetal2022} attribute the powering of the bump to the collision of the ejecta with a CSM density enhancement at $t \simeq 110 ~{\rm days}$, and at $R_{\rm EC} = t_{\rm bump} v_{\rm ex} \simeq 1.5 \times 10^{16}$. The similarity of the spectra at $t=83~{\rm days}$ and at $t=109~{\rm days}$ shows that the extra heating must affect both the photosphere and the H$\alpha$ emitting gas in a similar manner. 
For example, an increase in the H$\alpha$ emission due to the compression of the emitting region will not affect he photosphere, hence this cannot be an explanation to the bump.  

At $t \simeq 110 ~{\rm days}$ the photosphere is at $R_{\rm ph}=2.3 \times 10^{15} \cm\simeq 0.15R_{\rm EC}$ \citep{Franssonetal2022}. From geometrical considerations the fraction of the energy that the ejecta-CSM interaction radiates at $R_{\rm EC}$ that reaches the photosphere is (e.g., \citealt{ChevalierFransson1994}) 
\begin{eqnarray}
\begin{aligned} 
f_{\rm ph} & = \frac{1}{2} 
\left( 1-\sqrt{1-\frac{R^2{\rm ph}}{R^2_{\rm EC}}} \right) \simeq \frac{1}{4} \frac{R^2{\rm ph}}{R^2_{\rm EC}} 
\\ & = 0.0056 \left( \frac{R{\rm ph}}{0.15 R_{\rm EC}} \right)^2. 
\end{aligned}
\label{eq:fph}
\end{eqnarray}
The ejecta-CSM collision is very inefficient in powering the bump. The total dissipated energy in the collision should be 
\begin{eqnarray}
\begin{aligned} 
E_{\rm EC} & \simeq 10^{51} 
\left( \frac{E_{\rm rad,b}}{1.8 \times 10^{48} \erg} \right)
\left( \frac{f_{\rm EC,rad}}{0.5} \right)^{-1}
\\ & \times 
\left( \frac{f_{\rm ph,rad}}{0.5} \right)^{-1}
\left( \frac{f_{\rm ph}}{0.007} \right)^{-1} 
\erg, 
\end{aligned}
\label{eq:Eec}
\end{eqnarray}
where $f_{\rm CE,rad}$ is the fraction of energy that ends in radiation in the ejecta-CSM collision and $f_{\rm ph,rad}$ is the fraction of the radiation of the photosphere that ends in the visible (observed band). For example, some X-ray radiation from the collision region might simple be reflected by the photosphere and will not contribute to the increase in the visible luminosity. 

Although the amount of energy that equation (\ref{eq:Eec}) gives is possible to release in the collision, e.g., if the ejecta collides with a CSM extra mass of $\simeq 0.2 M_\odot$, it seems to be an extreme case because the collision should take place within 15 days. 

The larger problem is that this collision has no marks on the spectrum (because the spectra at $t=83~{\rm days}$ and at $t=109~{\rm days}$ are basically identical). 
The CSM and ejecta are shocked to X-ray emitting temperatures in such a collision of a relative velocity of $\ga 10^4 \km \s^{-1}$. However, to be efficient in radiating the energy in about 15 days the shell should cool to low temperatures where it emits in the visible. Otherwise, $f_{\rm EC,rad} \ll 0.5$ and the demand on the collision energy is much larger. Therefore, in such a case the cooling shell will affect the spectrum. 

My conclusion is that it is extremely unlikely that ejecta-CSM collision powered the bump at $t_{\rm bump} \simeq 110~{\rm days}$. 
I turn to a powering by jets from the central NS, either a newly born NS in a CCSN or an old NS in a CEJSN event.

\subsection{Powering the bump with jets} 
\label{subsec:BumpJets}

I use the toy model that \cite{KaplanSoker2020a} built to estimate the timescale and luminosity of a jet-driven bump in the light curve of CCSNe. I refer to this interaction in the last panel of Fig. \ref{Fig:schematic}.  

The toy model assumes that the two opposite jets that the central NS (or a BH) launches are active for a very short time. The toy model assumes then that the interaction of the jets with the ejecta lasts for a very short time. \cite{KaplanSoker2020a} term the short interaction of each of the two jets with the ejecta as an off-center `mini-explosion'. 
The jet-ejecta interaction shocks a zone around the interaction region, the `cocoon'
(one cocoon per each of the two jets). By a simple analytical derivation  \cite{KaplanSoker2020a} find the following expressions for the typical time scale (rise time) of the bump and for the extra luminosity of the bump 
\small
\begin{eqnarray}
\begin{aligned} 
& t_{\rm toy, b} = 56 
\left(\frac{\epsilon_E}{0.01}\right)^{-1/4}
\left(\frac{\epsilon_V}{0.067}\right)^{3/4}
\left(\frac{\kappa_{\rm c}}{0.38 \cm^2\g^{-1}}\right)^{1/2}\\ & \times
\left(\frac{E_{\rm SN}}{2\times 10^{51} \erg}\right)^{-1/4} 
\left(\frac{M_{\rm SN}}{10 M_{\odot}}\right)^{3/4}
 {~\rm d}, 
\end{aligned}
\label{eq:ttoy}
\end{eqnarray}
\normalsize
and
\small
\begin{eqnarray}
\begin{aligned} 
& L_{\rm toy,b} = 4.2\times 10^{41} 
\left(\frac{\epsilon_E}{0.01}\right) 
\left(\frac{\epsilon_V}{0.067}\right)^{-1}
\left(\frac{\sin{\alpha_{\rm j}}}{0.087}\right) \\ & \times
\left(\frac{\beta}{0.5}\right)
\left(\frac{\kappa_{\rm c}}{0.38  \cm^2\g^{-1}}\right)^{-1}
\left(\frac{t_{\rm j,0}}{100 {~\rm d}}\right) 
 \\ & \times
\left(\frac{E_{\rm SN}}{2\times 10^{51} \erg}\right)^{3/2} 
\left(\frac{M_{\rm SN}}{10 M_{\odot}}\right)^{-3/2}
\erg \s^{-1},  
\end{aligned}
\label{eq:Ltoy}
\end{eqnarray}
\normalsize
respectively. In the above expressions 
$\epsilon_E \equiv E_{\rm 1j}/E_{\rm SN}$, where $E_{\rm 1j}$ is the energy that one jet deposits into one cocoon, $E_{\rm SN}$ is the total energy of the ejecta (mainly kinetic energy at late times), $M_{\rm SN}$ is the total ejecta mass, 
$\epsilon_V \equiv M_{\rm 1co}/M_{\rm SN}$, where $M_{\rm 1co}$ is the mass in one cocoon, $\alpha_{\rm j}$ is the half opening angle of each jet, $\kappa_{\rm c}$ is the opacity, and $\beta$ is the ratio of the radius at which the `mini-explosion' occurs to the ejecta outer radius. For the jet-ejecta interaction time (the time of the mini-explosion) I scale in this study with $t_{\rm j,0}=100~{\rm days}$. 

According to the properties of the bump of SN~2019zrk (section \ref{subsec:Bump}) I also use the following constraints on some toy model variables. For the rise time of the bump and its total radiated extra energy I get 
\begin{equation}
t_{\rm toy,b} = 15  {~\rm d} 
\label{eq:tbump}
\end{equation}
and
\begin{equation}
\begin{aligned} 
L_{\rm toy,b} t_{\rm toy,b} = E_{\rm rad, b} \simeq 1.8 \times 10^{48} \erg, 
\end{aligned}
\label{eq:Eradbump}
\end{equation}
respectively. I assume that the front of the ejecta moves at a faster velocity than the observed velocity after ejecta-CSM collision of $v_{\rm ex} \simeq 1.6 \times 10^4 \km \s^{-1}$ (section \ref{sec:Explosion}), and take 
\begin{eqnarray}
\begin{aligned} 
E_{\rm SN} & = 0.3 M_{\rm SN} (2 \times 10^4 \km  \s^{-1} )^2 
\\ & = 
2.39 \times 10^{52} 
\left(\frac{M_{\rm SN}}{10 M_{\odot}}\right) \erg. 
\end{aligned}
\label{eq:EsnToy}
\end{eqnarray}
Since by the time of the bump the photosphere has already moved deep into the ejecta (section \ref{subsec:CSM}) I take a lower value of the radius of jet-ejecta interaction, i.e., I take $\beta=0.1$ instead of $\beta=0.5$. For the same reason I take a lower value of the mass of the ejecta that the jet interacts with  $\epsilon_V < 0.067$, i.e., I scale here with $\epsilon_V = 0.01$.
For lack of further information I leave the variables $\kappa_{\rm c}$ and $\sin{\alpha_{\rm j}}$ as in the scaling of \cite{KaplanSoker2020a}. 

I emphasize that the set of the toy model parameters is not unique. My goal is to demonstrate that a jet-powered bump can work for reasonable physical values. 

From equations (\ref{eq:ttoy}), (\ref{eq:tbump}) and (\ref{eq:EsnToy}) I find with the above value of $\beta$ and $\epsilon_V$
\small
\begin{eqnarray}
\begin{aligned} 
2 & \simeq 
\left(\frac{\epsilon_E}{0.01}\right)^{-1/4}
\left(\frac{\epsilon_V}{0.01}\right)^{3/4}
\\ & \times
\left(\frac{\kappa_{\rm c}}{0.38 \cm^2\g^{-1}}\right)^{1/2}
\left(\frac{M_{\rm SN}}{10 M_{\odot}}\right)^{1/2}
\end{aligned}
\label{eq:ttoyF}
\end{eqnarray}
\normalsize
while from equations (\ref{eq:Ltoy}) -  (\ref{eq:EsnToy}) I find 
\small
\begin{eqnarray}
\begin{aligned} 
0.06 & \simeq   
\left(\frac{\epsilon_E}{0.01}\right) 
\left(\frac{\epsilon_V}{0.01}\right)^{-1}
\left(\frac{\sin{\alpha_{\rm j}}}{0.087}\right) \\ & \times
\left(\frac{\beta}{0.1}\right)
\left(\frac{\kappa_{\rm c}}{0.38  \cm^2\g^{-1}}\right)^{-1}
\left(\frac{t_{\rm j,0}}{100 {~\rm d}}\right) .
\end{aligned}
\label{eq:LtoyF}
\end{eqnarray}
\normalsize

The solution of equations (\ref{eq:ttoyF}) and (\ref{eq:LtoyF}) for the last two parameters of the toy model reads 
 \small
\begin{eqnarray}
\begin{aligned} 
\epsilon_E \simeq 6 \times 10^{-4} 
\qquad {\rm and} \qquad 
M_{\rm SN} \simeq 10 M_\odot . 
\end{aligned}
\label{eq:FInalParam}
\end{eqnarray}
\normalsize
I emphasise again that this is not a unique solution, but rather presents a plausible set of values. 
For these values and using equation (\ref{eq:EsnToy})the two jets together carry an energy of 
\small
\begin{eqnarray}
\begin{aligned} 
E_{\rm 2j, b} \simeq 2.9 \times 10^{49} \erg \simeq 16 E_{\rm rad, b}. 
\end{aligned}
\label{eq:E2jets}
\end{eqnarray}
\normalsize
About $6 \%$ of the total energy of the two post-explosion jets that power the bump ends in the bump extra radiation.

The modelling procedure of bumps with mini-explosions assumes that the cocoon and jets do not breakout from the photosphere. Namely, the very hot cocoon stays inside the photosphere. This implies that the mini-explosion does not lead to UV/X-ray burst. The main effect is to heat somewhat the polar caps of the photosphere (along the jets' axis), and by that to increase the luminosity. Much more energetic jets might lead to strong shock-breakout and UV/X-ray burst. Namely, the hot cocoon reaches the photosphere. In that case the interaction is no longer a mini-explosion. On the other extreme of mini-explosions that are powered by weak and brief jets there are very strong jets that might last for a long time. Such very energetic jets might power peculiar superluminous supernova like SN 2018cow and similar transients, whether in CEJSN (e.g., \citealt{SokeretalGG2019, Metzger2022FBOT,  Soker2022FBOT}) or in CCSNe (e.g., \citealt{KashiyamaQuataert2015, Tsunaetal2021, Gottliebetal2022cow, Guarinietal2022}). 
  
My main conclusion from this entire section is that it is much more likely that jets powered the sharp bump in the light curve of SN~2019zrk at $t \simeq 110~{\rm days}$ than an ejecta-CSM interaction. 

\section{Summary} 
\label{sec:Summary}

I analysed some properties of the 2009ip-like transient SN~2019zrk \citep{Strotjohanntal2021, Franssonetal2022}, and  applied jets to account for its pre-explosion, explosion, and post-explosion powering (Fig. \ref{Fig:schematic}). 

In section \ref{sec:PreExplosion} I presented my view that an interaction with a binary companion is the most likely explanation to the ejection of a massive CSM. The energy source of the mass ejection is mainly the accretion onto a companion, which might be a main sequence star, a WR star, or a NS/BH, and jets carry the accretion energy to the RSG envelope and by that ejecting the massive CSM. 
As well, jets are most likely to have powered the pre-explosion outburst (pre-cursor), as with some other ILOTs (other names include gap objects; luminous red novae; intermediate luminosity red transients). 

In section \ref{sec:Explosion} I considered two models for the CSM. In the first the CSM is optically thick in the first $\simeq 45~{\rm days}$ as \cite{Franssonetal2022} suggested. I found that the total ejecta (explosion) energy should be $E_{\rm ej,tot} \simeq {\rm several} \times 10^{52} \erg$ (equation \ref{eq:Ef}). In the second model that I discussed here the CSM is of lower mass and it is optically thin from early times. The absence of wide hydrogen lines at early times is because the ejecta is hydrogen-poor. In that case the explosion energy is $E_{\rm ej,tot} \approx 10^{52} \erg$ (equation \ref{eq:Eejecta}). 
In both cases the explosion must be driven by jets because neutrino-driven explosion models cannot supply this energy. 

In section \ref{sec:PostExplosion} I analysed the powering of the short and bright (sharp) bump in the declining light curve of SN~2019zrk at $t_{\rm bump} \simeq 110~{\rm days}$. I concluded that ejecta-CSM interaction is extremely unlikely to power such a sharp bump (section \ref{subsec:CSM}). Instead, I used a toy model from the literature to present one set of parameters (not unique) for jets that might power such a bump (section \ref{subsec:BumpJets}). I concluded that late (post-explosion) jets are the most likely explanation for the bump of SN~2019zrk, and possibly sharp bumps in the declining lightcurves of other CCSNe, like the two bumps in the lightcruve of SN~2019tsf that \cite{Zenatietal2022} study in a recent paper. Late sources of accreted mass that might lead to late jet-launching include fallback after the explosion (e.g., \citealt{DellaValleetal2006, Moriyaetal2010, AkashiSoker2022, Pellegrinoetal2022}), feeding the young NS (or BH) by an inflated main sequence companion (e.g., \citealt{Ogataetal2021, Hoberetal2022}), and the feeding of a pre-existing NS/BH by the CCSN ejecta (e.g., \citealt{Fryeretal2014, Becerraetal2015, Becerraetal2019, AkashiSoker2020}). At this time I leave all possibilities open for SN~2019zrk, as it is not clear whether it was a CCSN event or a CEJSN event.  
 
The powering of both the pre-explosion outburst, which is similar to outbursts of some ILOTs (luminous red novae;  intermediate-luminosity red transients), and late bumps by jets might be the case also in ILOTs.
Some ILOTs are known to be powered by jets because of their bipolar morphology, e.g.,  Nova 1670 (CK Vulpeculae; \citealt{Kaminskietal2021}). I therefore raise the possibility that the sharp bump in the light curve of the luminous red novae (ILOT) AT~2021biy at about 330 to 400 days post maximum \citep{Caietal2022} was powered by jets.     

On a broader scope, this study further connects peculiar transient events, specifically here 2009ip-like transient events, with CCSNe by further arguing that jets drive all these events, from regular CCSNe through superluminous CCSNe and to many other peculiar and super-energetic transient events (section \ref{sec:intro}; see \citealt{Soker2022Rev} for a review).  
In recent years a number of research groups presented the view that jets power not only energetic and peculiar CCSNe, but also regular CCSNe (e.g., \citealt{Soker2010, Papishetal2015, Izzoetal2019, Piranetal2019}). In the majority of the CCSNe that do not have pre-collapse rapid rotation the accretion disk that launches the jets is a stochastic intermittent accretion disk that launches jittering jets (e.g., \citealt{PapishSoker2011, ShishkinSoker2022}). The source of the stochastic angular momentum is the pre-collapse convection motion in the core (e.g.,  e.g., \citealt{GilkisSoker2014, Quataertetal2019, ShishkinSoker2021}).
 
This study strengthens the call to include jet-powering in the analysis of all luminous transient events of massive stars (excluding pair instability supernova; \citealt{RakavyShaviv1967}), whether CCSNe or CEJSNe, and most likely in analysing most CCSNe even if are not luminous.

\section*{Acknowledgments}

I thank Amit Kashi and an anonymous referee for useful comments. This research was supported by a grant from the Israel Science Foundation (769/20).

\section*{Data availability}
The data underlying this article will be shared on reasonable request to the corresponding author.  

\end{document}